# Seizures and epilepsy after intracerebral hemorrhage: an update

Laurent Derex [1, 2, #], MD, PhD; Sylvain Rheims [3,4,5], MD, PhD, Laure Peter-Derex [4, 5, 6,#], MD, PhD

[1] Stroke Center, Department of Neurology, Neurological Hospital, Hospices Civils de Lyon, University of Lyon, France

[2] University Claude Bernard Lyon 1, Research on Healthcare Performance (RESHAPE), INSERM U1290, Lyon, France

[3] Department of Functional Neurology and Epileptology, Hospices Civils de Lyon, University of Lyon, France

[4] Lyon 1 University, France

[5] Lyon Neuroscience Research Center, INSERM U1028 - CNRS UMR 5292, Lyon, France

[6] Center for Sleep Medicine and Respiratory Disease, Croix-Rousse Hospital, Hospices Civils de Lyon, University of Lyon, France

#Corresponding authors:

Dr. Laurent Derex, Stroke Center, Department of Neurology, Neurological Hospital, Hospices Civils de Lyon, 59 boulevard Pinel, 69677 Bron cedex, France.
ORCID: 0000-0002-0909-8900
E-mail: laurent.derex@chu-lyon.fr

Dr. Laure Peter-Derex: Center for Sleep Medicine and Respiratory Disease, Croix-Rousse Hospital, Hospices Civils de Lyon, University of Lyon, 103 Grande rue de la Croix-Rousse, 69004, Lyon, France.
ORCID: 0000-0002-9938-9639
E-mail: laure.peter-derex@chu-lyon.fr

## Abstract

Seizures are common after intracerebral hemorrhage, occurring in 6 to 15% of the patients, mostly in the first 72 hours. Their incidence reaches 30% when subclinical or non-convulsive seizures are diagnosed by continuous electroencephalogram. Several risk factors for seizures have been described including cortical location of intracerebral hemorrhage, presence of intraventricular hemorrhage, total hemorrhage volume, and history of alcohol abuse. Seizures after intracerebral hemorrhage may theoretically be harmful as they can lead to sudden blood pressure fluctuations, increase intracranial pressure and neuronal injury due to increased metabolic demand. Some recent studies suggest that acute symptomatic seizures (occurring within seven days of stroke) are associated with worse functional outcome and increased risk of death despite accounting for other known prognostic factors such as age and baseline hemorrhage volume. However, the impact of seizures on prognosis is still debated and it remains unclear if treating or preventing seizures might lead to improved clinical outcome. Thus, the currently available scientific evidence does not support the routine use of antiseizure medication as primary prevention among patients with intracerebral hemorrhage. Only prospective adequately powered randomized controlled trials will be able to answer whether seizure prophylaxis in the acute or longer term settings is beneficial or not in patients with intracerebral hemorrhage.

# Declaration

**Funding**

Not applicable

**Conflicts of interest**

Laurent Derex has nothing to disclose

Sylvain Rheims has nothing to disclose

Laure Peter-Derex is the principal investigator of the PEACH trial (Clinicaltrials.gov, NCT 02631759)

**Availability of data and material**

Not applicable

**Code availability**

Not applicable

**Ethics approval**

Not applicable

**Consent to participate**

Not applicable

**Consent for publication**

Not applicable

# Introduction

Intracerebral hemorrhage (ICH) accounts for 10-15 % of all strokes and results in death or severe disability in more than 60 % of patients [1,2]. The acute phase of an ICH is often complicated by seizures, likely reflecting the disruptive effect on neuronal networks of the hematoma and surrounding edema [3]. Survivors of acute ICH are also at high risk for long-term sequelae, including late post-stroke epilepsy [4]. The goal of this update is to summarize the available literature, focusing on the epidemiology, diagnosis, electrophysiological features and treatment of ICH-related seizures and epilepsy, and to highlight the areas needing further research.

# Terminology

Seizures manifesting as a consequence of brain injuries such as ICH are usually separated into acute symptomatic seizures (ASS) and unprovoked seizures (US) depending on the time point of occurrence [5]. The International League Against Epilepsy (ILAE) defines ASS as seizure occurring within seven days of stroke, while seizures are unprovoked if they manifest after more than one week [6]. Previously, ASS have been referred to as 'early seizures' and US as 'late seizures', but in the last years, these terms have been abandoned. US are then further classified as recurrent (if patients previously experienced an early seizure) or delayed [3].

If at least one US occurs in a patient with a structural lesion such as ICH increasing the risk of further seizures, the probability of further seizures is similar to the general recurrence risk after two unprovoked seizures (at least 60%), leading to the diagnosis of epilepsy according to the ILAE definition of epilepsy [7] (**Figure 1**). Thus, a single US due to stroke should be considered as post-stroke epilepsy [5].

Some authors have hypothesized that delayed seizures are associated with different risk factors as compared to recurrent seizures. ASS in the acute phase of ICH could be primarily caused by mechanical effects of the expanding hemorrhage, the disruption of cortical networks by the hematoma via its structural damaging properties and/or irritation of the cortex due to products of blood metabolism. In contrast, seizures manifesting for the first-time in a delayed manner may be attributable to the more subtle cortical damaging effects of underlying cerebral small vessel disease, acting slowly but progressively over time [4] or may be caused by cortical irritation from hemosiderin depositions and gliotic scarring as well as inflammatory processes involved in epileptogenesis [8,9] (**Figure 2**).

# Epidemiology of ICH-related seizures

## Seizures at the acute phase of ICH

Seizures are more common in hemorrhagic than ischemic stroke [10,3] but the reported incidence of ICH-related ASS is highly variable. Comparisons between studies are difficult because of different patient populations, seizure criteria, and follow-up periods. In prospective studies, ASS ranged from 5% to 14% in patients with ICH [11-13] as compared to 5 to 6% in patients with ischemic stroke [14,15]. The majority of ICH-related ASS occurs within the first 72 hours supporting the recommendations to monitor patients in stroke or intensive care units during the acute phase [15-19]. Studies of ICH patients using continuous electroencephalography monitoring (cEEG) in the intensive care unit have reported substantially higher rates of subclinical seizures [10,20]. A study reported electrographic seizures, i.e. seizures without any clinically detected symptoms, in 28% of 63 patients with ICH [10] (**Figure 1**). In this series, cEEG detected four times as many electrographic seizures as occurred clinically. Another study of 102 consecutive patients with ICH who underwent cEEG showed that seizures occurred in one third of patients [20]. Convulsive seizures occurred prior to cEEG in 19%, another 18% had electrographic seizures, and 5% had both convulsive seizures preceding the cEEG and electrographic seizures during the monitoring. This study only included a critically ill subpopulation of patients with ICH who underwent cEEG and therefore likely overestimates the frequency of seizures in a general ICH population. Only one of the 18 patients with electrographic seizures also had a recognized clinical seizure while on cEEG [20]. In patients with electrographic seizures, the first seizure was detected within the first hour of cEEG monitoring in 56% and within 48 hours in 94%. This series identified proximity to the cortical surface as a predictor of electrographic seizures, corroborating prior reports that found lobar more likely than deep hemorrhages to cause clinical and subclinical seizures [10]. Electrographic seizures were twice as common (33% vs 15%) in patients with expanding hemorrhages (an increase in ICH volume of 30% or more between admission and 24-hour follow-up CT scan). In the light of these studies, using cEEG to enhance detection rates of seizures may represent a valuable approach to gather valid information on true incidence rates of acute seizures associated with ICH [21].

## Delayed seizures after ICH

Regarding the rate of US, in a retrospective study of 615 primary ICH patients who survived for longer than 3 months, 83 (13.5%) developed post-stroke epilepsy [22]. The risk of new-onset post-stroke epilepsy was highest during the first year after ICH with cumulative incidence of 6.8%. Other studies have reported that during two years of follow-up, 8% to 10% of ICH survivors develop additional US [15,23]. Another study with longer follow-up showed a cumulative risk of US of 11.8% five years after ICH [24]. In a prospective cohort of consecutive adults with spontaneous ICH [23], the presence of lobar brain microbleeds (especially if ≥3) was associated with the risk of US, pointing to a potential link with the underlying vasculopathy (cerebral amyloid angiopathy). US were also associated with a worse functional outcome after 3 years of follow-up, suggesting that US may either have a direct influence on outcome or may simply reflect the severity of the underlying disease.

**Risk factors of ICH-related seizures**

Several risk factors for ASS and/or US have been described, including cortical or subcortical location of ICH, presence of intraventricular hemorrhage, total hemorrhage volume, history of alcohol abuse and surgical hematoma evacuation [25,15,26,4,27,21,28,29]. Retrospective analysis of the observational Helsinki ICH Study showed that US occurred more frequently in younger patients, with larger ICH, when the ICH involves the cortex, and after ASS [24]. Differences in the clinical manifestation of epilepsy in elderly and younger adults can lead to underestimation of epilepsy incidence in older people. Convulsive seizures may become less frequent, and clinical seizure manifestations may be more difficult to recognize in the elderly [24]. Larger ICH volume leads to more extensive neuronal damage with higher risk of epileptogenesis and US [28,29]. Another study showed that subcortical hematoma location and ASS increased the risk of post-stroke epilepsy after primary ICH in long-term survivors, while hypertension seemed to reduce the risk, likely because ICHs are more commonly localized to the deeper structures in hypertensive patients [22]. Regarding the association of ASS with post-stroke epilepsy, it has been suggested that early epileptiform activity could increase the metabolic demand causing secondary brain damage and gliotic scarring [24]. However, other studies have concluded that ASS did not predict the risk of developing US [25,23].

A clinical score for late seizure risk prediction after ICH has been proposed [24]. The CAVE score (0–4 points) was created to estimate the risk of US in individual patients, with 1 point for each of cortical involvement, age <65 years, volume >10 mL, and early seizures within 7 days of ICH. As these 4 variables are readily available soon after the ICH, the score is easy to calculate and shows an almost linear risk increase. The risk of US was 0.6%, 3.6%, 9.8%, 34.8%, and 46.2% for scores 0 to 4, respectively. It is estimated that only 15% of patients with ICH have the highest risk scores of 3 to 4; even in these patients, the risk is <50% for several years. The score has been validated in an independent prospective ICH cohort. However, compared with the derivation cohort c-statistic of 0.81 (0.76–0.86), the validation cohort c-statistic was relatively low at 0.69 and had a confidence interval from 0.59 to 0.78. Further validation in other cohorts appears warranted in order to establish generalizability.

Some authors have also pointed out that the suboptimal score's predictive performance in the validation cohort raises the possibility of a more complex biological substrate for US after ICH [4]. The analysis of a single-centre longitudinal cohort study of ICH identified largely different risk factors for delayed seizures following ICH when compared to recurrent seizure events in patients with a known history of seizure in the acute ICH phase [4]. Delayed seizures were strongly associated with known clinical, neuroimaging or genetic risk factors for cerebral small vessel disease. On the contrary, acute ICH characteristics (increasing ICH volume and severity of neurologic deficit at onset) were predictors of recurrent seizure risk. This study has shown that availability of genetic (APOE genotype) and MRI data (presence of exclusively lobar cerebral microbleeds) may substantially improve ability to stratify risk for late seizures.

# Electrophysiological patterns in ICH

The diagnosis of seizures in the context of ICH relies on clinical symptoms and EEG recordings including routine EEG and/or cEEG monitoring.

## Patterns of EEG activity recorded in ICH

Several patterns of abnormal EEG activity related to the presence of a focal brain injury (such as ICH) have been reported. Non-epileptic patterns consist in background abnormalities such as focal to diffuse slowing of EEG activity. Interictal epileptic activity include sporadic epileptiform discharges and periodic or rhythmic patterns [30]. According to their topography, periodic discharges (PDs) and rhythmic delta activity (RDA) are classified as lateralized (LPDs, LRDA), generalized (GPDs, GRDA) or bilateral independent (BIPDs) [30]. Ictal patterns, usually recorded in the context of cEEG monitoring, mainly present as evolving discharges of any type that reach a frequency > 4c/s; however, periodic EEG patterns that are time-locked to patients movements are also considered as ictal [31]. Among interictal patterns, LPDs are of particular interest, as they are considered to represent an ictal-interictal continuum state and are associated with a high prevalence of non-convulsive seizures [32]. The pathophysiology of LPDs is unclear; they may be the manifestation of an abnormal neuronal response in a localized cortical area potentially resulting from lesional of functional denervation [33]. They may also have a deleterious effect per se, as they are associated with worse outcome even in the absence of underlying detectable radiological brain lesion [34].

## Prognosis value of electrographic ictal and interictal patterns

The prognosis value of ictal and interictal patterns recorded on cEEG in the context of ICH remains poorly known as most data rely on retrospective studies. From a physiological point of view, invasive multimodality monitoring in comatose patients with spontaneous subarachnoid hemorrhage has shown that seizures recorded with intracortical electrodes are associated with elevated heart rate, blood pressure and respiratory rate [35]. Using the same type of recordings, it was reported that high frequency PDs are associated with brain oxygen level decrease without sufficient compensatory increase in cerebral blood flow, thus potentially leading to additional brain damage [36]. Vespa et al. reported that non-convulsive and convulsive seizure during the initial 72h after admission, including a majority of cEEG-detected seizures, were associated in ICH with worse neurologic function and brain edema as assessed by increased midline shift [10]. A close rate of cEEG detected seizures (around one third of ICH patients) was found in Claassen's study; in this work, PEDs were more frequently seen in hemorrhages closer to the cortex and were independently associated with poor outcome [20]. A recent clinical study found that the presence of epileptiform abnormalities on cEEG in acute ischemic stroke was independently associated with poor functional outcome, with a dose-dependent relationship [37]. However, these two latter studies being retrospective, most patients had severe conditions justifying cEEG, with a majority of poor outcomes.

Regarding the risk of late epilepsy, a prospective study demonstrated, in the context of ischemic stroke, that background activity asymmetry and interictal epileptiform discharges recorded on an EEG performed during the first 72h after admission were independent predictors of post-stroke epilepsy [38]. This study confirmed retrospective works showing that highly epileptic findings (electrographic seizures and LPDs) on cEEG in acute brain insult including ICH are associated with further development of new onset epilepsy (HR = 7.7 (95% CI:2.9-20.7) for LPDs alone and 11.4 (95% CI: 4-31.4) for LPDs and electrographic seizures) [39,40]. Such abnormalities could represent early neurophysiological biomarkers of epileptogenesis processes [41]. Thus, detecting electrographic ictal and "high risk" interictal patterns is crucial regarding prognosis including risk of epilepsy. Moreover, other cEEG patterns such as the presence of physiological sleep features or topographical organization of electrophysiological activity have been associated with better functional and mortality outcomes in ICH [42].

**Recordings and recommendations**

Several studies have highlighted the fact that a high number of seizures are not clinically detected in ICH, especially non-convulsive seizures in patients with altered consciousness and neurological symptoms secondary to the hemorrhagic lesion [20,10]. Routine EEG seldom allow to record seizures, and does not always demonstrate interictal discharges whereas cEEG allows to detect a higher number of seizures. It is worth mentioning that even cEEG sensitivity is far from perfect, as several seizures remain blind to scalp EEG and are only recorded using intracortical recordings performed as part of research protocols in comatose patients [43,35].

In practice, routine EEG is recommended in ICH patients with unexplained and persistent altered consciousness or in case of clinically suspected seizures [44]. However, less then 50% of ICH patients in neurointensive care units who fulfill guideline criteria benefit from routine EEG [45]. According to the American Clinical Neurophysiology Society recommendations, cEEG should be performed in order to identify non-convulsive seizures or non-convulsive status epilepticus in acute supratentorial brain injury with altered mental status but also in patient with PDs on routine EEG or in case of clinical paroxysmal events suspected to be seizures, for at least 24 to 48h [46]. In critically ill patients including ICH, cEEG use was recently showed to be associated with reduced in-hospital mortality [47]. In spite of these recommendations, cEEG remains under-used in stroke units, even in neurovascular intensive care settings. As the presence of interictal epileptiform discharges on standard EEG is predictive for the occurrence of ictal patterns on cEEG, routine EEG may help to select patients in whom cEEG is required [48]. Early recordings seem warranted as most seizures occur within the first 48h after admission for ICH [20]. New devices such as dry cap electrode EEG may allow to widen the utilization of cEEG in stroke patients [49].

# Antiseizure medication in ICH

## Acute phase management

*ICH-related ASS and prognosis*

Whether or not patients should receive antiseizure drugs (ASDs) as primary pharmacological prevention of seizures after spontaneous ICH remains a matter of debate [50,51,5].

Seizures after ICH may theoretically be harmful: they can lead to sudden blood pressure fluctuations, increase intracranial pressure and neuronal injury due to increased metabolic demand, and are independently associated with worse outcome in some series [10,18,13]. A cohort study including 5027 consecutive patients with acute ischemic or hemorrhagic stroke showed that patients with seizures occurring during inpatient stay had a higher mortality at 30-day and at 1-year post-stroke, longer hospitalization, and greater disability at discharge [52]. In another large study of 2325 patients with ICH, early seizures (≤7 days) were associated with worse functional outcome and increased risk of death at three months [13]. On the other hand, some studies have shown no association of seizures with early neurologic deterioration or mortality at 30 days or 1 year in patients with ICH [15,53]. In other observational studies, clinical seizures did not worsen long term outcome from ICH [54-56,28,11]. Moreover, the association observed between ASS and poor prognosis may only reflect underlying common factors to both outcomes such as large volumes of ICH.

*ASD prophylaxis and outcome*

The uncertainty about the relative risks and benefits of primary prevention of seizures translates into a wide variation in rates of prescribing ASDs after ICH, with up to 30% of physicians reporting routine use of these agents [18]. In a retrospective study conducted in two academic US centers, 98 (19.4%) out of 506 patients with primary ICH were started on prophylactic anticonvulsants [57]. Levetiracetam (97%) was most commonly prescribed. Age, lobar location, higher initial National Institutes of Health Stroke Scale (NIHSS) score, craniotomy, and prior ICH were independently associated with prophylactic anticonvulsant initiation. Prophylactic anticonvulsants were very commonly continued through hospital discharge and, in some cases, months or even years afterward.

ASDs may have associated toxicity and side effects such as fever, liver abnormalities, and cognitive dysfunction, depending on the specific medication. Some ASDs such as phenytoin and phenobarbital could also inhibit neural plasticity and hinder recovery [58]. Newer ASDs such as levetiracetam and lacosamide are better tolerated, have less drug-drug interactions and better side effect profiles, and show potential neuroprotective effect [59]. Regarding the occurrence of post-stroke epilepsy, no clinical trial has demonstrated that temporary ASD treatment after brain injury including stroke prevents or mitigates epileptogenesis [60-62]. Some retrospective studies [63,64] have suggested increased complications rate and worse outcome in ICH patients treated with ASDs while others have shown no association between ASD treatment and epilepsy, disability, or death [65]; Battey *et al*., 2012). The design and the results of the studies published up to now are summarized in the **Table**. In an observational study of patients with acute ICH, the early use of ASDs was associated with severe disability and death, independent of other significant predictors of poor outcome [63]. However, most of the patients in this cohort received phenytoin and these results may not be generalizable to other ASDs. Interestingly, a prospective study of 98 patients, of whom 40 received prophylactic ASDs, found that phenytoin was associated with poor outcome at 3 months but that levetiracetam was not [64]. Another study showed that after adjustment for multiple factors associated with poor outcome, prophylactic levetiracetam was not associated with worse functional outcome at 3 months [66]. Other studies comparing levetiracetam and phenytoin in patients with ICH have suggested that levetiracetam was associated with improved cognitive outcomes at discharge and fewer seizures [67] as well as improved long-term outcome [68]. Another retrospective analysis of a cohort of patients with acute ICH showed no association between ASD treatment and mortality or outcome at 3 months [69]. The authors concluded that any detected association could arise by confounding by indication, in which the most severely affected patients are those in whom ASDs are prescribed. These results are in line with those of a more recent retrospective study which again showed no association between prophylactic ASD treatment and worse functional outcome at discharge or at one year [57]. The single randomized, double-blinded, placebo-controlled trial of ASD for seizure prevention in ICH was limited by a small sample size (n=72) and the use of clinically reported events without the use of cEEG [62]. In this single-center trial comparing immediate valproic acid for 1 month with placebo, a nonsignificant decrease in early seizures (1/36 versus 4/36; P=0.4) was noted but no effect was observed on further seizures during a follow-up of one year. According to a recent systematic review and meta-analysis of 7 studies with a total of 3241 patients, the use of ASDs as primary prevention among adult patients with spontaneous ICH is not associated with improved neurological function nor with decreased incident clinically evident seizures during long-term follow-up [70]. However, most studies included in this analysis were observational studies with unclear risk of bias and randomized controlled trials (RCTs) are lacking. Moreover, significant heterogeneity was observed across studies in the duration of patient follow-up, and studies with only short time horizons may have failed to detect some clinically important seizures. There was also a high variability in the definition of early and late seizures, and most studies did not utilize continuous electroencephalographic monitoring.

In the light of the currently available data, clinical guidelines recommend against the use of prophylactic antiseizure medication in patients with acute ICH [71,51,5]. According to the European Stroke Organisation (ESO) guidelines for the management of post-stroke seizures and epilepsy, clinicians may decide individually to temporarily administer primary ASD prophylaxis (for not longer than the acute phase) in some subgroups of patients with ICH, e.g. in those with cortical involvement [5]. The American Heart Association/American Stroke Association guidelines for the management of spontaneous ICH recommend the use of ASDs for patients with either clinical seizures or electroencephalographic evidence of seizures with decreased mental status [51]. In the absence of adequately powered RCTs, evidence for all these recommendations is very low.

*Treatment of acute symptomatic seizures*

In absence of evidence-based relation between ASS and long-term risk of post-ICH epilepsy, and as highlighted by the current guidelines, there is no indication of initiating antiseizure medication in patients with ICH-related ASS. On the other hand, after a first ASS, it might be important to transiently reduce the risk of seizure-related complications in the early post-ICH period, including the risk of fall, injuries and aspiration pneumonia, especially in the elderly. If the physician decides to initiate an ASD, the treatment choice should primarily take into account the pharmacokinetics characteristics of the drug, with a preference for an ASD that can be titrated very quickly, administered intravenously, and which lacks significant drug-drug interactions. The two most commonly prescribed ASDs that meet these characteristics are levetiracetam (LEV) and lacosamide (LCS).

Furthermore, the specific situation of repetitive ASS should be considered. Seizure cluster, which is usually defined as occurrence of > 3 seizures in 24 hours, is significantly associated with risk of developing status epilepticus [72]. Early status epilepticus occurs in about 1% of all patients with stroke, but in 27% of patients with ASS [56]. In addition, risk of early status epilepticus is two-fold greater in patients with ICH than in those with ischemic stroke [56]. The issue of status epilepticus is particularly important in the elderly population because its mortality is age-dependent, lowest in the young and highest in the elderly [73]. Cerebrovascular diseases represent half of the acute symptomatic causes of status epilepticus after 60 years [74]. Overall, the principles of pharmacological management of cluster of ASS or status epilepticus should not differ from the current guidelines with first line therapy relying on acute administration of benzodiazepines [75]. In patients with benzodiazepine-refractory status epilepticus, levetiracetam, sodium valproate and fosphenytoin can be considered, without difference between them in efficacy or safety outcomes, even in older adults [76].

*Treatment of electrographic epileptic activity*

Treatment indication of PDs using ASDs remains a matter of debate; it has been proposed that treatment should be considered for PDs > 2Hz and/or associated with faster frequencies and/or of sharply contoured morphology, as they have the greatest seizure predictive value or may be more damaging [77,36]. Clinical assessment using benzodiazepine trial may be useful in therapeutic decision, as well as associated neuroimaging signs of neuronal injury potentially secondary to excitotoxicity, such as cortical hyperintensities in diffusion-weighted MRI [78,79]. The impact of curative or prophylactic ASD treatment of PDs or electrographic seizures on long term outcome remains uncertain. In practice, most patients with epileptic findings on cEEG are treated with ASD, which often remain prescribed over long time periods and may lead to underestimation of "true" epilepsy incidence in these patients [40].

**Management of delayed seizures**

US after ICH occurs relatively commonly and usually necessitates secondary prophylaxis [80]. Due to their considerable social consequences such as driving and working limitations and negative impact on quality of life, prevention of seizure recurrence is of utmost importance in patients with ICH. US recurrence risk is reported to be higher than 70% in 10 years [81]. Patients who develop US after ICH run a particularly high risk of seizure recurrence, and if antiseizure medication is not started after US, more than 90% of patients can expect further seizures [82,83,15].

Although the benefit of secondary ASD prophylaxis has not been proven in RCTs, guidelines state that this therapy needs to be considered in patients with ICH after one US [5]. Two RCTs compared efficacy of two different ASDs after stroke. In these underpowered trials, seizure freedom rates after 12 months did not differ between levetiracetam and carbamazepine [84] and between lamotrigine and carbamazepine [85]. The choice of the ASD should take into account patient profile, with particular attention to the enzymatic inducing effect and the potential atherogenic role of carbamazepine [86]. If secondary ASD prophylaxis is employed, it may be continued permanently, as seizure recurrence risk after ASD withdrawal in patients with lesional epilepsy has been reported to be higher than 50% [87,88,5].

## Conclusions and future directions

Seizures are a common complication of acute ICH, particularly in subcortical and cortical hemorrhages which carry higher risk of developing seizures compared to hemorrhages in deeper structures. The majority of seizures which occur after hospital admission in ICH patients is purely electrographic and can only be diagnosed with cEEG monitoring. Most electrographic seizures are detected within the first 48 hours of monitoring. The frequency and the impact on prognosis of clinical seizures, clinically unrecognized electrographic seizures and PEDs in acute ICH patients remain unclear. Further research should evaluate the input of cEEG on the therapeutic management of ICH in large prospective studies. Current evidence for management of post-stroke seizures and epilepsy is very low. It remains unclear if treating or preventing seizures might lead to improved clinical outcome after ICH. Future studies should focus on the preventive use of newer ASDs among patients at high risk of both seizures (according to the CAVE score for instance) and poor outcome. The ongoing pilot randomized placebo-controlled PEACH trial evaluates the potential efficacy of levetiracetam in acute ICH patients with cEEG monitoring (Clinicaltrials.gov, NCT 02631759). Only prospective adequately powered RCTs will be able to answer whether seizure prophylaxis in the acute or longer term settings is beneficial or not in ICH patients.

# Table caption

**Table. Studies evaluating antiseizure drugs after intracerebral hemorrhage**

ASDs indicates antiseizure drugs

The % indicates the rate of patients treated with each ASD

# Figure caption

## Figure 1

**Box: definitions of clinical epileptic seizure, epilepsy and electrographic seizures**

**Clinical epileptic seizure:** transient occurrence of signs and/or symptoms due to abnormal excessive or synchronous neuronal activity in the brain [89].

**Epilepsy:** disorder of the brain characterized by an enduring predisposition to generate epileptic seizures, and by the neurobiological, cognitive, psychological and social consequences of this condition. The definition of epilepsy requires the occurrence of at least one epileptic seizure [89]. *Practical definition*: 1) at least two unprovoked (or reflex) seizures occurring > 24h apart 2) one unprovoked (or reflex) seizure and a probability of further seizures similar to the general recurrence risk (at least 60%) after two unprovoked seizures, occurring over the next 10 years 3) diagnosis of an epilepsy syndrome [7].

**Electrographic seizures:** rhythmic discharge or spike and wave pattern with definite evolution in frequency, location or morphology lasting at least ten seconds [90]. These seizures refer to "subclinical" seizures, i.e. ictal discharges without detected clinical signs and symptoms either because the neurological state of the patient may not allow for the observation of additional seizure symptoms or because potential symptoms related to the epileptic discharge cannot be detected with a routine neurological examination.

## Figure 2

**Classification and proposed mechanisms of seizures and epilepsy following intracerebral hemorrhage**

47. Bermeo-Ovalle A (2019) Bringing EEG Back to the Future: Use of cEEG in Neurocritical Care. Epilepsy Curr 19 (4):243-245. doi:10.1177/1535759719858350

48. Koren J, Herta J, Draschtak S, Potzl G, Pirker S, Furbass F, Hartmann M, Kluge T, Baumgartner C (2015) Prediction of rhythmic and periodic EEG patterns and seizures on continuous EEG with early epileptiform discharges. Epilepsy Behav 49:286-289. doi:10.1016/j.yebeh.2015.04.044

49. Doerrfuss JI, Kilic T, Ahmadi M, Weber JE, Holtkamp M (2020) Predictive value of acute EEG measurements for seizures and epilepsy after stroke using a dry cap electrode EEG system - Study design and proof of concept. Epilepsy Behav 104 (Pt B):106486. doi:10.1016/j.yebeh.2019.106486

50. Morgenstern LB, Hemphill JC, 3rd, Anderson C, Becker K, Broderick JP, Connolly ES, Jr., Greenberg SM, Huang JN, MacDonald RL, Messe SR, Mitchell PH, Selim M, Tamargo RJ, American Heart Association Stroke C, Council on Cardiovascular N (2010) Guidelines for the management of spontaneous intracerebral hemorrhage: a guideline for healthcare professionals from the American Heart Association/American Stroke Association. Stroke 41 (9):2108-2129. doi:10.1161/STR.0b013e3181ec611b

51. Hemphill JC, 3rd, Greenberg SM, Anderson CS, Becker K, Bendok BR, Cushman M, Fung GL, Goldstein JN, Macdonald RL, Mitchell PH, Scott PA, Selim MH, Woo D, American Heart Association Stroke C, Council on C, Stroke N, Council on Clinical C (2015) Guidelines for the Management of Spontaneous Intracerebral Hemorrhage: A Guideline for Healthcare Professionals From the American Heart Association/American Stroke Association. Stroke 46 (7):2032-2060. doi:10.1161/STR.0000000000000069

52. Burneo JG, Fang J, Saposnik G, Investigators of the Registry of the Canadian Stroke N (2010) Impact of seizures on morbidity and mortality after stroke: a Canadian multi-centre cohort study. Eur J Neurol 17 (1):52-58. doi:10.1111/j.1468-1331.2009.02739.x

53. Leira R, Davalos A, Silva Y, Gil-Peralta A, Tejada J, Garcia M, Castillo J, Stroke Project CDGotSNS (2004) Early neurologic deterioration in intracerebral hemorrhage: predictors and associated factors. Neurology 63 (3):461-467. doi:10.1212/01.wnl.0000133204.81153.ac

54. Kilpatrick CJ, Davis SM, Tress BM, Rossiter SC, Hopper JL, Vandendriesen ML (1990) Epileptic seizures in acute stroke. Arch Neurol 47 (2):157-160. doi:10.1001/archneur.1990.00530020053014

55. Burn J, Dennis M, Bamford J, Sandercock P, Wade D, Warlow C (1997) Epileptic seizures after a first stroke: the Oxfordshire Community Stroke Project. BMJ 315 (7122):1582-1587. doi:10.1136/bmj.315.7122.1582

56. Labovitz DL, Hauser WA, Sacco RL (2001) Prevalence and predictors of early seizure and status epilepticus after first stroke. Neurology 57 (2):200-206. doi:10.1212/wnl.57.2.200

57. Mackey J, Blatsioris AD, Moser EAS, Carter RJL, Saha C, Stevenson A, Hulin AL, O'Neill DP, Cohen-Gadol AA, Leipzig TJ, Williams LS (2017) Prophylactic Anticonvulsants in Intracerebral Hemorrhage. Neurocrit Care 27 (2):220-228. doi:10.1007/s12028-017-0385-8

58. Troisi E, Paolucci S, Silvestrini M, Matteis M, Vernieri F, Grasso MG, Caltagirone C (2002) Prognostic factors in stroke rehabilitation: the possible role of pharmacological treatment. Acta Neurol Scand 105 (2):100-106. doi:10.1034/j.1600-0404.2002.1o052.x

(2014) European Stroke Organisation (ESO) guidelines for the management of spontaneous intracerebral hemorrhage. Int J Stroke 9 (7):840-855. doi:10.1111/ijs.12309

72. Haut SR (2015) Seizure clusters: characteristics and treatment. Curr Opin Neurol 28 (2):143-150. doi:10.1097/WCO.0000000000000177

73. Leppik IE (2018) Status epilepticus in the elderly. Epilepsia 59 Suppl 2:140-143. doi:10.1111/epi.14497

74. Rohracher A, Reiter DP, Brigo F, Kalss G, Thomschewski A, Novak H, Zerbs A, Dobesberger J, Akhundova A, Hofler J, Kuchukhidze G, Leitinger M, Trinka E (2016) Status epilepticus in the elderly-A retrospective study on 120 patients. Epilepsy Res 127:317-323. doi:10.1016/j.eplepsyres.2016.08.016

75. Glauser T, Shinnar S, Gloss D, Alldredge B, Arya R, Bainbridge J, Bare M, Bleck T, Dodson WE, Garrity L, Jagoda A, Lowenstein D, Pellock J, Riviello J, Sloan E, Treiman DM (2016) Evidence-Based Guideline: Treatment of Convulsive Status Epilepticus in Children and Adults: Report of the Guideline Committee of the American Epilepsy Society. Epilepsy Curr 16 (1):48-61. doi:10.5698/1535-7597-16.1.48

76. Chamberlain JM, Kapur J, Shinnar S, Elm J, Holsti M, Babcock L, Rogers A, Barsan W, Cloyd J, Lowenstein D, Bleck TP, Conwit R, Meinzer C, Cock H, Fountain NB, Underwood E, Connor JT, Silbergleit R, Neurological Emergencies Treatment T, Pediatric Emergency Care Applied Research Network i (2020) Efficacy of levetiracetam, fosphenytoin, and valproate for established status epilepticus by age group (ESETT): a double-blind, responsive-adaptive, randomised controlled trial. Lancet 395 (10231):1217-1224. doi:10.1016/S0140-6736(20)30611-5

77. Bauerschmidt A, Rubinos C, Claassen J (2018) Approach to Managing Periodic Discharges. J Clin Neurophysiol 35 (4):309-313. doi:10.1097/WNP.0000000000000464

78. Lansberg MG, O'Brien MW, Norbash AM, Moseley ME, Morrell M, Albers GW (1999) MRI abnormalities associated with partial status epilepticus. Neurology 52 (5):1021-1027. doi:10.1212/wnl.52.5.1021

79. Canas N, Breia P, Soares P, Saraiva P, Calado S, Jordao C, Vale J (2010) The electroclinical-imagiological spectrum and long-term outcome of transient periictal MRI abnormalities. Epilepsy Res 91 (2-3):240-252. doi:10.1016/j.eplepsyres.2010.07.019

80. Balami JS, Buchan AM (2012) Complications of intracerebral haemorrhage. Lancet Neurol 11 (1):101-118. doi:10.1016/S1474-4422(11)70264-2

81. Hesdorffer DC, Benn EK, Cascino GD, Hauser WA (2009) Is a first acute symptomatic seizure epilepsy? Mortality and risk for recurrent seizure. Epilepsia 50 (5):1102-1108. doi:10.1111/j.1528-1167.2008.01945.x

82. Berger AR, Lipton RB, Lesser ML, Lantos G, Portenoy RK (1988) Early seizures following intracerebral hemorrhage: implications for therapy. Neurology 38 (9):1363-1365. doi:10.1212/wnl.38.9.1363

83. Sung CY, Chu NS (1989) Epileptic seizures in intracerebral haemorrhage. J Neurol Neurosurg Psychiatry 52 (11):1273-1276. doi:10.1136/jnnp.52.11.1273

84. Consoli D, Bosco D, Postorino P, Galati F, Plastino M, Perticoni GF, Ottonello GA, Passarella B, Ricci S, Neri G, Toni D, Study E (2012) Levetiracetam versus carbamazepine in patients with late poststroke seizures: a

| Study (Year) | Design | Total sample size | ASDs | Outcome |
|---|---|---|---|---|
| **Messé et al. (2009)[63]** | Prospective cohort | 295 | Phenytoin 78% | ASDs associated with disability and death at 3 months |
| **Naidech et al. (2009) [64]** | Prospective cohort | 98 | Phenytoin Levetiracetam | Phenytoin associated with poor outcome at 3 months |
| **Szaflarski et al. (2010) [68]** | Randomized comparative trial | 52 | Phenytoin Levetiracetam | Levetiracetam associated with improved long term outcome |
| **Gilad et al. (2011) [62]** | Randomized placebo-controlled trial | 72 | Valproic acid 100% | Valproic acid associated with non-significant decrease in early seizures |
| **Reddig et al. (2011) [65]** | Retrospective cohort | 157 | Phenytoin 57% | ASDs not associated with in-hospital death |
| **Taylor et al. (2011) [67]** | Retrospective cohort | 269 | Phenytoin 29% Levetiracetam 71% | Levetiracetam associated with improved cognitive outcome and decrease in seizures |
| **Battey et al. (2012) [69]** | Retrospective cohort | 1 182 | Phenytoin 68% Levetiracetam 30% | ASDs not associated with death at 3 months |
| **Sheth et al. (2015) [66]** | Retrospective cohort | 744 | Levetiracetam 86% | Levetiracetam not associated with outcome at 3 months |
| **Mackey et al. (2017) [57]** | Retrospective cohort | 506 | Levetiracetam 97% | ASDs not associated with long term outcome |

**Table. Studies evaluating antiseizure drugs after intracerebral hemorrhage**
ASDs indicates antiseizure drugs
The % indicates the rate of patients treated with each ASD

**Figure 1**

**Box: definitions of clinical epileptic seizure, epilepsy and electrographic seizures**

**Clinical epileptic seizure:** transient occurrence of signs and/or symptoms due to abnormal excessive or synchronous neuronal activity in the brain [89].

**Epilepsy:** disorder of the brain characterized by an enduring predisposition to generate epileptic seizures, and by the neurobiological, cognitive, psychological and social consequences of this condition. The definition of epilepsy requires the occurrence of at least one epileptic seizure [89]. *Practical definition*: 1) at least two unprovoked (or reflex) seizures occurring > 24h apart 2) one unprovoked (or reflex) seizure and a probability of further seizures similar to the general recurrence risk (at least 60%) after two unprovoked seizures, occurring over the next 10 years 3) diagnosis of an epilepsy syndrome [7].

**Electrographic seizures:** rhythmic discharge or spike and wave pattern with definite evolution in frequency, location or morphology lasting at least ten seconds [90]. These seizures refer to "subclinical" seizures, i.e. ictal discharges without detected clinical signs and symptoms either because the neurological state of the patient may not allow for the observation of additional seizure symptoms or because potential symptoms related to the epileptic discharge cannot be detected with a routine neurological examination.

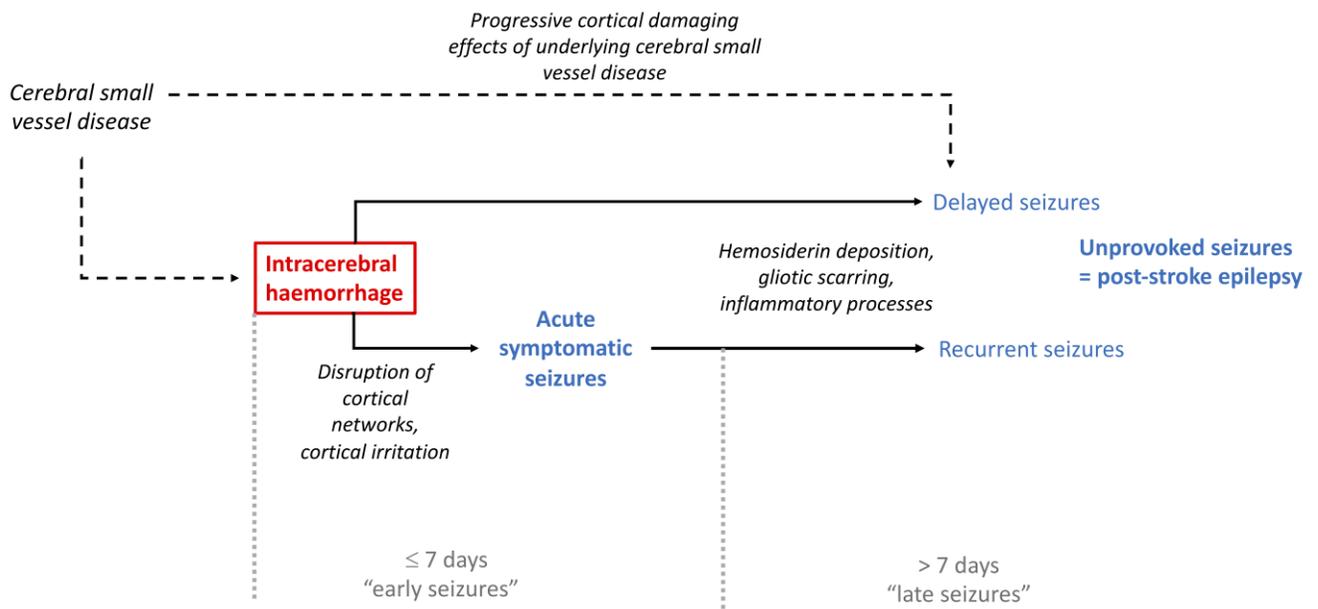